\documentclass[fleqn,10pt]{wlscirep}
\usepackage[utf8]{inputenc}
\usepackage[T1]{fontenc}
\usepackage{array}
\usepackage{graphicx}
\usepackage{float}
\DeclareUnicodeCharacter{2212}{-}

\title{Machine-learned metrics for predicting the likelihood of success in materials discovery}

\author[1, *]{Yoolhee Kim}
\author[1]{Edward Kim}
\author[1]{Erin Antono}
\author[1]{Bryce Meredig}
\author[1]{Julia Ling}

\affil[1]{Citrine Informatics, Redwood City, 94063 CA, USA}
\affil[*]{ykim@citrine.io}


\begin{abstract}
Materials discovery is often compared to the challenge of finding a needle in a haystack. While much work has focused on accurately predicting the properties of candidate materials with machine learning (ML), which amounts to evaluating whether a given candidate is a piece of straw or a needle, less attention has been paid to a critical question: Are we searching in the right haystack? We refer to the haystack as the \emph{design space} for a particular materials discovery problem (i.e. the set of possible candidate materials to synthesize), and thus frame this question as one of \emph{design space selection}. In this paper, we introduce two metrics, the Predicted Fraction of Improved Candidates (PFIC), and the Cumulative Maximum Likelihood of Improvement (CMLI), which we demonstrate can identify discovery-rich and discovery-poor design spaces, respectively. Using CMLI and PFIC together to identify optimal design spaces can significantly accelerate ML-driven materials discovery. 

\end{abstract}
\begin{document}

\flushbottom
\maketitle
\thispagestyle{empty}

\section*{INTRODUCTION}
\par A significant challenge in materials discovery is the vast, often untenable, space of potential experiments that could be performed in any given materials optimization effort. Desirable, novel materials exist as needles in enormous proverbial haystacks. Brute force searches of these haystacks, which may represent material compositions, crystal structures, or synthesis parameters, are prohibitively expensive and time-consuming. Therefore, efficient methods for discovering needles are sought to reduce the number of experiments required to discover novel materials that meet a given set of performance specifications  \cite{meredig2014combinatorial,isayev2015materials,rajan2015materials,ramprasad2017machine}. 
\par Computational techniques may alleviate this challenge by screening promising candidate materials via predicting whether each candidate from the haystack exceeds some threshold performance (e.g. thermoelectric figure of merit). This haystack may represent a large, high-dimensional design space \cite{gomez2016design, greeley2011computational,isayev2015materials,meredig2014combinatorial, kim2017virtual, borboudakis2017chemically}, or the set of all relevant materials whose performances are unknown.
\par Model accuracy is a well-studied factor in determining whether a materials discovery project using computational techniques will be successful \cite{snoek2012practical, ward2018matminer, ward2016general}. Model accuracy has therefore been widely used as a success metric for machine-learned structure-property relations. A less widely considered factor is the quality of the haystack, or the design space. Depending on the quality of the design space, discovering an improved material may range from trivially easy (if all untested materials are superior) to impossible (if all untested materials are inferior). Jia et al.\cite{jia2019anthropogenic} and Kauwe et al.\cite{kauwe2019can} have shown, respectively, that random and ML-guided materials search strategies may achieve success depending on the underlying difficulty - or quality - of the design space. Thus, there is a critical need for a quantitative method that determines the quality of the design space, which strongly impacts materials discovery success. In other words, we desire a method for finding the best haystacks in which to search for needles. 
\par To the best of the authors' knowledge, there is no current work on quantifying the quality of a given design space, where a \emph{high quality} space yields materials discovery success with fewer required experiments \cite{sanchez2018inverse}. While recent studies have begun to explore the effect of choosing different training and test sets for discovery on the practicality of discovering new and useful materials, broadly quantifying these effects in terms of successful materials discoveries remains unexplored \cite{kauwe2019can, meredig2018can, raccuglia2016machine, jia2019anthropogenic}.
\par In this work, we first demonstrate how design space quality is a critical factor in materials development success. We illustrate this by benchmarking sequential learning success versus design space quality through simulated sequential learning on existing materials data sets. Second, we introduce a procedure to initialize the training data and design space which better reflects how known and unknown materials data are distributed \cite{kauwe2019can, meredig2018can,ward2018matminer}. Finally, we present a design space evaluation system using two novel metrics - the PFIC and CMLI scores - for quantifying the quality of a given design space. This design space evaluation system gives insight on the likelihood of success in a given materials development project, and enables data-driven selection between different materials discovery projects.

\section*{METHODS}

\begin{figure}[ht]
\centering
\includegraphics[width=\linewidth]{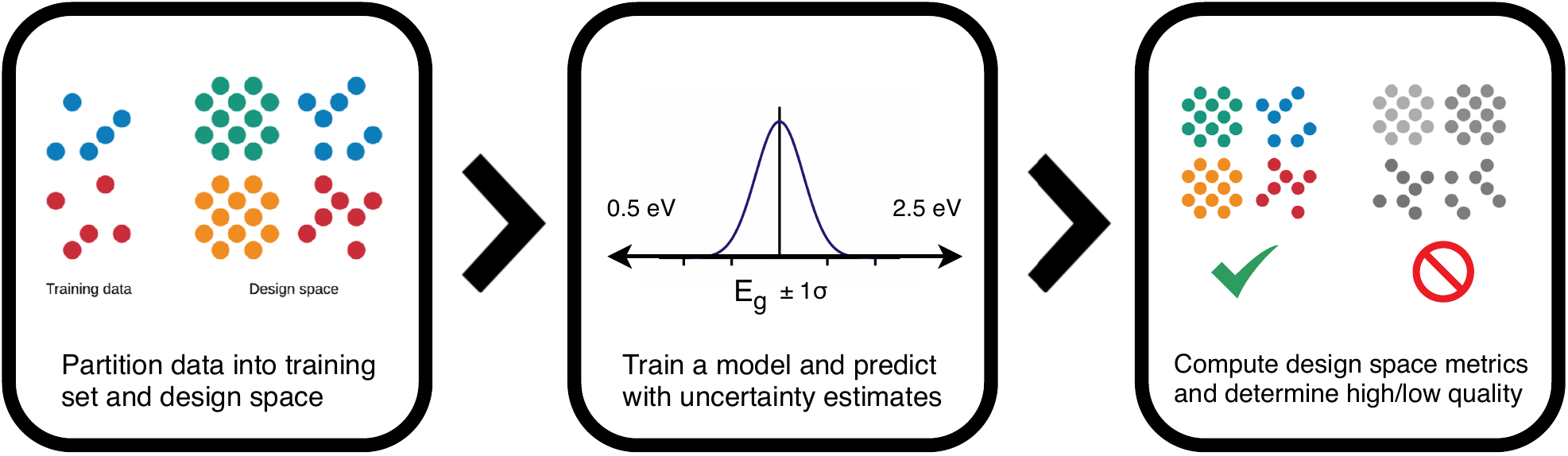}
\caption{Overall schematic of the methods presented in this work. Data are first initialized into a training set and a design space. This is followed by training a model using the training data, and predicting material performances across the design space candidates. Finally, predictive design space metrics are computed and used to evaluate high quality and low quality design spaces.}
\label{fig:overall_schematic}
\end{figure}

The methods presented in this work are highlighted in Figure \ref{fig:overall_schematic}. Using a novel data partitioning scheme to best represent the nature of in-lab materials discovery, a machine-learned model is trained on a set of training data. Following this, predictions for material properties (e.g., band gap) are computed along with uncertainty estimates across materials in the held-out design space. Finally, these predictions and uncertainties are used to produce \emph{design space metrics} that are indicative of the difficulty of materials discovery for a particular design space. Or, in other words, these metrics evaluate how difficult it is to find a needle in a given haystack.

\subsection*{Sequential learning simulations}
\par To enable the rapid computation of quantitative materials discovery success and design space quality metrics across a broad variety of materials categories, as well as different splits of the training set and design space, \emph{in silico} simulations of sequential learning were used to avoid the overhead of in-lab experimental trials. \emph{In silico} simulations of sequential learning reduce the time required to run experiments to effectively zero. 
\par Sequential learning was simulated using two publicly available datasets - the Materials Project and Harvard Clean Energy Project datasets, shown in Table \ref{table:benchmarks}. Across a number of trials, we split each dataset into a training set and design space to assess the quality of the design space on the overall likelihood of sequential learning success. The candidates in the design spaces held out from model training were treated as an approximation to infinite design spaces commonly encountered during in-lab materials development. The methodology of splitting these benchmark datasets into a training set and design space is discussed in the following section. 
\par The training set was used to train machine learning models, then the trained models were used to produce predictions with uncertainties for each of the design space candidates' material properties of interest. At each sequential learning iteration, the candidate with the highest performance was then selected to ``measure,'' at which point the true performance of that candidate was revealed to the model, and that candidate was added to the training set. Prior to the next sequential learning iteration, the model was retrained on the training data, including the added design space candidate. Various metrics were recorded at each iteration of sequential learning for analysis, and these metrics are discussed in the following section.
\par Machine learning models were built for five properties in total. The objective was to maximize, minimize, or tune the value of the target property, depending on the case. Table \ref{table:benchmarks} lists the datasets and properties used for simulating sequential learning. A variety of benchmark datasets were chosen to represent different materials classes, as well as both computational and experimental datasets.  The machine learning model inputs were representative of the degrees of freedom that could be adjusted in a laboratory setting, such as the composition and the processing parameters of a material. The output of the machine learning model was the material property to be optimized, such as the band gap.
\par In this work, the open-source \emph{lolo} random forest library was used for the machine learning algorithm \cite{lolo}. Random forests make predictions for a new point based on the values of training data points that are nearby in the input space. Given the clustered nature of many materials datasets, this approach to making predictions make random forests particularly well-suited to materials applications \cite{Ling2017}. However, we stress that the methods introduced in this work are model-agnostic and are thus compatible with a wide variety of modern algorithms, including Gaussian processes and neural networks.
\par 20 identical sequential learning trials were performed for each training set and design space in order to capture the variance due to stochastic model training. To analyze the effect of design space quality on sequential learning success, each trial of the sequential learning process was performed for 50 iterations of sequential learning (where each iteration represents the addition of one new data point to the training set and the retraining of the machine learning model), or until a design space candidate that performed better than the best training data candidate was found. The maximum likelihood of improvement (MLI) and the maximum expected improvement (MEI) were used as acquisition functions to determine the performance of the design space candidates \cite{Ling2017}. 50 to 100 different training data and design space initializations were tested per dataset. Each training data and design space split exhibited a varying level of sequential learning difficulty by virtue of the training-to-design-space ratio and fraction of improved candidates in the design space.
\par Therefore, sequential learning trials were run in order to analyze the relationship between design space quality and success in a materials discovery project without the need for in-lab experiments.

\begin{center}
\begin{table} [H]
\caption{Benchmark datasets used in sequential learning simulations and analysis of predictive design space metrics.}
\begin{tabular}{p{4.4cm}p{1.5cm}rp{4.1cm}p{3.7cm}}
 Dataset & Origin  & Dataset Size & Output Property & Objective \\
 \hline\hline
Harvard Clean Energy Project\cite{hachmann2011harvard} & DFT  & 30,000  & Power conversion efficiency & Maximize  \\
 & &  & HOMO Energy & Maximize \\ \hline
Materials Project\cite{jain2013commentary} & DFT & 30,000  & Formation energy & Minimize \\
 & &  & Band gap & Tune to 1.5 eV \\
 & &  & Total magnetization & Maximize absolute value \\ \hline
Melting Points\cite{Bradley2014} & Experiment & 3041  & Melting point temperature & Maximize \\ \hline
Superconductors\cite{SuperconductorDataset} & Experiment & 585 & Superconducting $T_c$ & Maximize \\ \hline
UCSB Thermoelectrics\cite{gaultois2013data} & Experiment & 1093 & Seebeck coefficient & Maximize \\
 & &  & Electrical resistivity & Minimize \\ \hline
Strehlow \& Cook\cite{strehlow1973compilation} & Experiment & 1449 & Band gap & Tune to 1.5 eV
\end{tabular}
\label{table:benchmarks}
\end{table}
\end{center}

\subsection*{Data initialization}
\par While randomly-chosen splits of training and test data are common throughout machine learning \cite{ward2016general, ward2018matminer}, recent works by Meredig et al. and Kauwe et al. \cite{meredig2018can,kauwe2019can} have shown that the choice of training and test data significantly influence estimates of generalization error in machine-learned models, along with the success of materials discovery.
\par
\begin{figure}[ht]
\centering
\includegraphics[width=\linewidth]{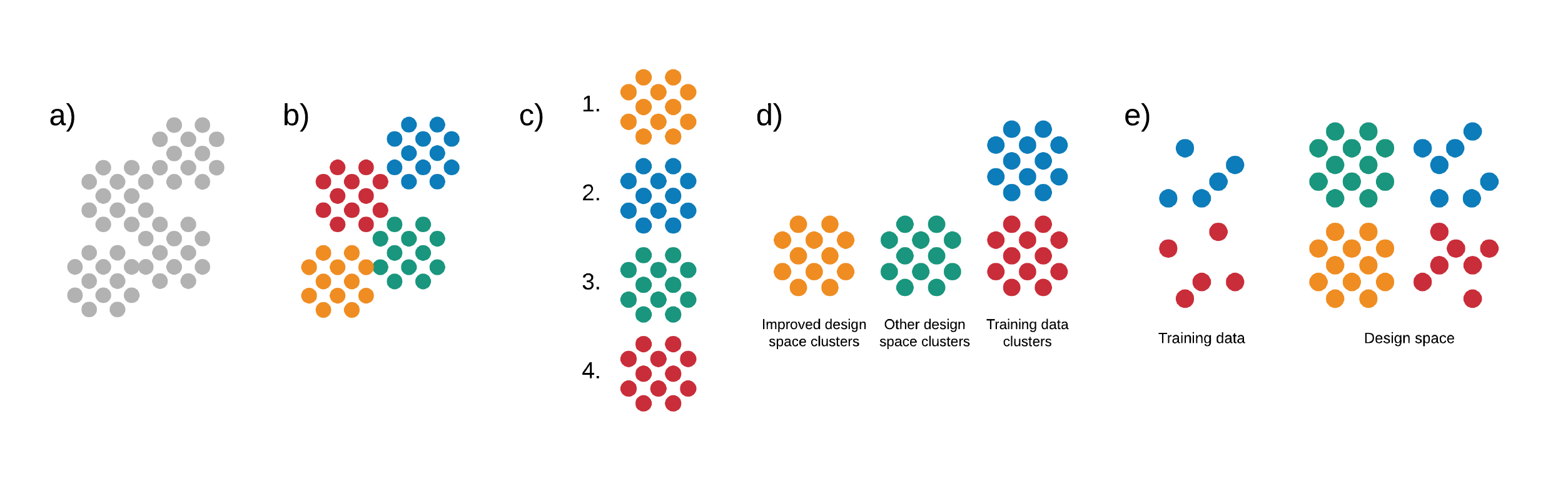}
\caption{Diagram of the data initialization approach. Data initialization begins with a) a given dataset. The data initializer b) clusters the full dataset, then c) ranks the clusters based on the best performing candidate within. After ranking the clusters, d) the clusters are assigned as either training clusters or design clusters. Finally, the data initializer e) subsamples the training clusters so that some elements from the training clusters are included in the design space.}
\label{fig:data_initializer}
\end{figure}

 To reflect the clustered nature of materials data \cite{meredig2018can, kauwe2019can}, which are generally derived from human-biased historical results \cite{jia2019anthropogenic}, a cluster-based data initialization method was developed and used in this study. This data initialization method better matched experimental materials discovery projects, since experiments are typically run over similar types of materials, resulting in clustered training data \cite{jia2019anthropogenic}. Figure \ref{fig:data_initializer} shows a diagram of this data initialization method.
\par We also tested random selection of training sets and design spaces, which often yielded trivially solvable sequential learning simulations when initial training data included at least one material nearby to improved, held-out materials. Machine-learned algorithms were able to successfully discover an improved material within five iterations of sequential learning in such design spaces in almost all trials. Existing experimental and computational sequential learning studies have typically explored dozens to hundreds of iterations \cite{granda2018controlling, brandt2017rapid} before discovering novel or improved materials, suggesting that these splits of training and test data were unrealistic, and confirming that a more complex data initialization method was required. 
\par The data initialization method was performed via the following algorithm:

\begin{enumerate}
    \item Data were first clustered into $n$ clusters. The number of clusters, $n$, in this work was tested at 16, 32, 64, and 128.
    \item The $n$ clusters were ranked based on the best performing candidate within each cluster, with respect to the material property of interest.
    \item These clusters were divided into training and design space clusters. Different trials varied the fraction of improved clusters that were assigned as design clusters, thereby varying the quality of the design space.  
    \item Training clusters were split into two parts.  Some of the data were included in the training set, and others were included in the design space.  This approach was adopted because typical design spaces include both interpolative (similar to training data) and extrapolative (different from training data) candidates. No data points were included in both the training set and design space.
\end{enumerate}
\par The data initialization algorithm exposed several parameters that were used to modulate the difficulty of the sequential learning problem. These parameters are discussed in more detail in the Supplementary Information. A higher fraction of improved clusters in the design space (i.e. more needles in the haystack) corresponds to an easier materials discovery problem. Additionally, the difficulty of the problem could be influenced by which points from a training cluster were included in the design space.  If the best points in a given training cluster were assigned to the design space, then the materials discovery problem was found to be substantially easier because improved candidates existed near to the training data.  On the other hand, if no improved candidates from training clusters were included in the design space, then it was a more difficult sequential learning problem because extrapolation was required to identify improved candidates. 
\par This clustering approach to divide the training set and design space was developed to use in our sequential learning simulations, as well as to calculate predictive design space metrics. While this data initialization approach may seem complex, it was developed with great care to realistically reflect how known and unknown data are distributed in materials datasets \cite{meredig2018can,kauwe2019can,ward2018matminer,jia2019anthropogenic}.

\section*{RESULTS AND DISCUSSION}
\subsection*{Effect of design space quality on sequential learning success}
\par Sequential learning is a workflow that requires iterative testing of materials suggested by a machine learning model. It would be extremely valuable to a researcher to know from the outset of a materials discovery project whether the project is likely to take 10 or 100 experiments to find an improved material. For this reason, this paper quantifies sequential learning success as the number of iterations required to find an improved candidate in the design space (i.e. the number of draws from the haystack until a needle is found). This paper quantifies design space quality as the Fraction of Improved Candidates (FIC), or the fraction of candidates in the design space that perform better than the best training candidate (i.e. the fraction of the haystack that are needles). In practice, design space quality is unknown when searching for an undiscovered material. Therefore, predictive metrics correlated with the FIC are useful to assess design space quality.
\par In order to determine the effect of design space quality on sequential learning success, sequential learning simulations were run using the Materials Project and Harvard Clean Energy Project datasets from Table \ref{table:benchmarks}. These datasets were sub-sampled and divided into a training set and a design space using the data initialization method described in the previous section. The five properties used for these simulations, as well as their objectives, are listed in Table \ref{table:benchmarks}.
\par
\begin{figure}[ht]
\centering
\includegraphics[width=0.6\textwidth]{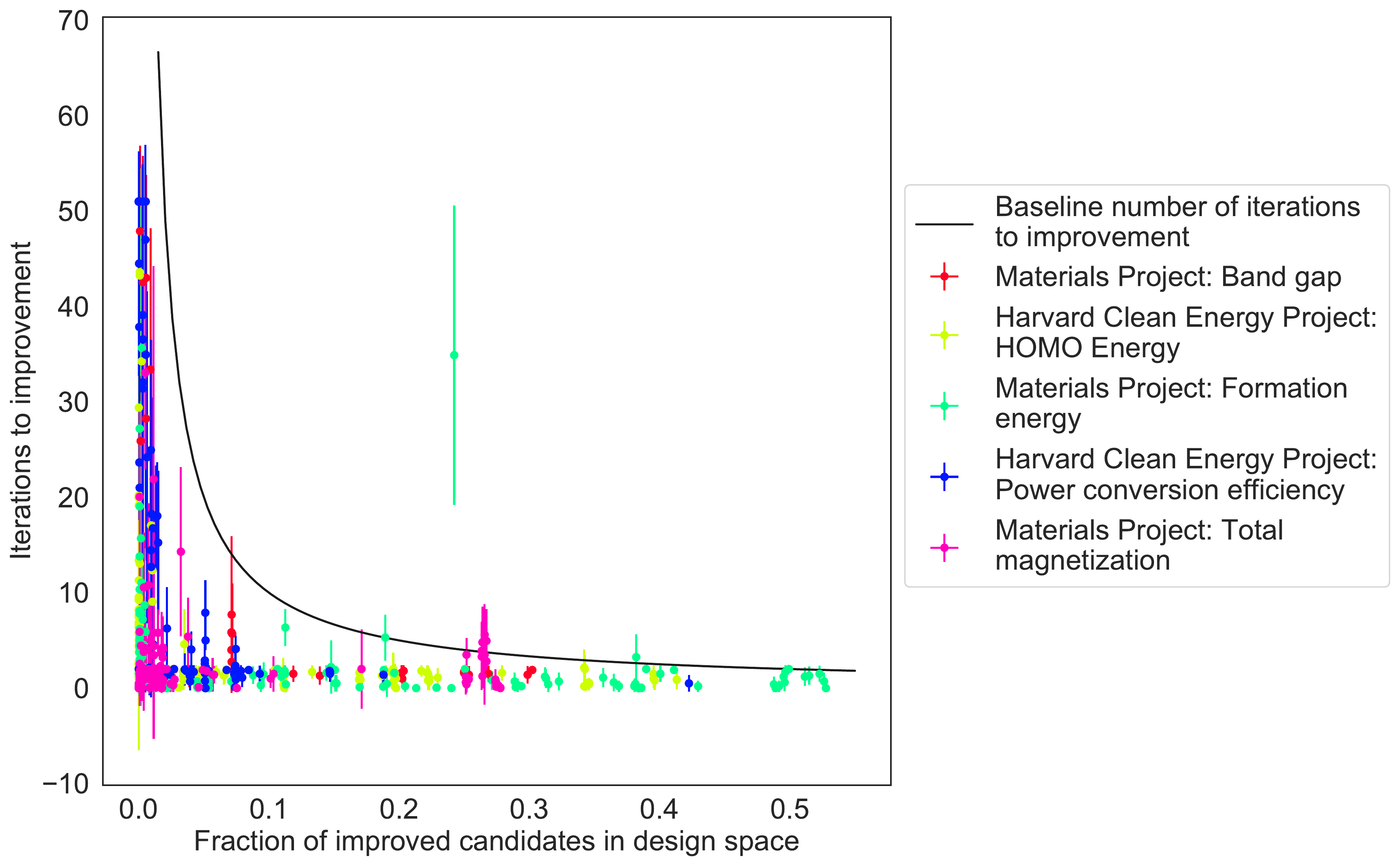}
\caption{Number of iterations required to find an improved candidate versus the fraction of improved candidates in the design space. Sequential learning simulations were run over five different benchmark simulations, represented by the different colors. Each point represents a different design space and training data initialization, and the error bars on each point represents the variance in number of iterations it takes to find an improved material over 20 sequential learning simulations with the same data initialization but stochastically trained random forest models. The black line represents the expected number of iterations to find an improved candidate via random search strategy, $1/p$.}
\label{fig:iters_to_improvement}
\end{figure}

 Figure \ref{fig:iters_to_improvement} shows sequential learning success, or the number of iterations required to find an improved candidate from the design space, as a function of the FIC, or the fraction of improved candidates in the design space, over these benchmark test cases. Each color in this scatter plot corresponds to a different sequential learning objective - maximizing, minimizing, or tuning of some target property from a benchmark dataset. Each data point in this scatter plot represents a separate data intialization of the training data and design space, and the error bars represents the standard deviation over 20 trials of identical data initializations caused by stochastic model training.
\par This figure shows that the number of iterations required to find an improved candidate is highly sensitive to the design space quality, or the FIC. For design spaces with low FIC, many iterations are required on average to find an improved candidate, and the number of iterations required has high variance. For design spaces with high FIC, a comparatively lower number of iterations are required to find an improved candidate. This result emphasizes the contribution of the design space quality to the difficulty of a sequential learning problem. Therefore, design space quality is expected to have a strong impact on the success of sequential learning projects, and materials discovery projects in general. Additional experiments were performed to analyze the relationship between other metrics, such as model quality, and the required number of iterations until an improvement is found. The results of these studies indicate that these other metrics, unlike design space quality, are not as strongly tied to sequential learning success. The details of these tests along with their results are available in the Supplementary Information. 
\par Additionally, Figure \ref{fig:iters_to_improvement} highlights how sequential learning nearly always outperforms the baseline random-selection strategy, which is shown by the black curve. Assuming design space candidates are chosen randomly at each sequential learning iteration, the required number of iterations to find an improved candidate can be approximated as the expected number of trials until success for a Bernoulli process. This expected number of trials is given by $1/p$, where $p$ is the fixed per-trial probability of success, which is equal to the FIC. This baseline curve is an approximation and is only accurate for large design spaces where sampling without replacement does not significantly affect the probability of success $p$, or the probability for finding an improved candidate.
\par Figure \ref{fig:iters_to_improvement} also shows a clear outlier corresponding to the formation energy from the Materials Project dataset \cite{jain2013commentary}. Upon investigating this outlier case, it was found that the number of iterations to improvement shows a bimodal distribution. In some trials, the model is able to find an improvement immediately, in the first iteration of sequential learning. In other trials, the iterations to improvement is much larger, between 21 and 46 iterations. Based on this evidence, our hypothesis is that there are several poor-performance candidates near the improved candidates in the design space. The model may get ``lucky'' and discover an improved material immediately, or get ``unlucky'' and discover the bad candidate, misleading the model to not continue exploring that region near the other improved candidates. This outlier highlights the existence, and importance, of antagonistic data-splits in the validation and application of machine learning models for materials discovery. Future work includes further investigation of this effect and determining how to mitigate this outcome in experimental (non-simulated) sequential learning applications. 
\par In summary, the fraction of improved candidates in the design space (FIC) was identified as a key metric that is strongly correlated to how many experiments are required to find an improved material (i.e. the number of draws from the haystack until a needle is found), emphasizing the contribution of design space quality to the difficulty of a sequential learning project. The FIC can therefore be used to determine how likely a materials discovery project is to succeed.  However, we emphasize that in general, we do not know the FIC for a given design space \textit{a priori}. Thus, we would like to identify other metrics we can calculate based only on the initial training data that are predictive of the FIC.

\subsection*{Predictive metrics to evaluate design space quality}
\par Given the strong relationship between design space quality and sequential learning success (the number of iterations required to find an improved candidate), design space quality can be used as a key indicator to determine the difficulty of a sequential learning problem. However during in-lab materials discovery, design space quality, or the FIC (fraction of improved candidates in the design space), is not known. Therefore, predictive design space metrics to assess the design space quality are desirable.
\par Several predictive design space metrics were examined in this work. Each metric leveraged machine-learned predictions, and some of these metrics leveraged uncertainty estimates. These metrics were then compared against the FIC, the true design space quality. These metrics were calculated for 10 properties over six different benchmark datasets from Table \ref{table:benchmarks} which represent different materials classes, as well as both computational and experimental datasets. Each dataset had 50 to 100 different training data and design space initializations, where each split had varying FIC, or design space quality. Critically, sequential learning iterations were not required to calculate these predictive design space metrics and the FIC, since these quantities can be computed without any iterative data acquisition. Table \ref{table:tested_metrics} presents the Pearson correlations of the examined predictive design space metrics against the true design space quality, or the FIC. Additional details regarding the testing of these predictive metrics, along with their explicit definitions, can be found in the Supplementary Information.

\begin{table} [H]
\centering
\caption{Correlations of predictive metrics with true design space quality. Predictive metrics were tested across the datasets in Table \ref{table:benchmarks} with 10 trials for each data initialization. Metrics in boldface yielded the highest correlations with true design space quality.}
\begin{tabular}{r|c}
 Predictive Metric & Pearson Correlation with True Design Space Quality (FIC)\\
 \hline \hline
 \textbf{Predicted fraction of improved candidates (PFIC)} & \textbf{0.38} \\
 PFIC minus 1$\sigma$ uncertainty & 0.20 \\
 \textbf{Top-10 Cumulative MLI (CMLI)} & \textbf{0.31} \\
 Top-5 CMLI & 0.27 \\
 Predicted fraction of improved, interpolated candidates & 0.11 \\
 Ratio of extrapolated to all design candidates & 0.29 \\
\end{tabular}
\label{table:tested_metrics}
\end{table}

\par As shown in Table \ref{table:tested_metrics}, the PFIC and top-10 CMLI score achieved the highest correlations with true design space quality. Consequently, these two predictive metrics are further discussed in the following sections of this work. 

\subsubsection*{Predicted Fraction of Improved Candidates (PFIC) score}
\par The PFIC score is defined as the fraction of design space candidates that are predicted by the model to have improved performance over the best training data point. A machine learning model was fit to the training data to calculate the PFIC score. In this work, a random forest \cite{lolo} was used, where at each leaf node in this random forest, a linear model was fit to the training data. The number of trees was set to the training set size, the maximum tree depth was set to 30, the minimum number of samples required to split an internal node was set to 20, and regularization of the linear model coefficients was implemented. 
\par After training this random forest, the model then predicted the performance of all of the candidates in the design space. The PFIC score can then be calculated by:
\begin{equation} \label{PFIC_eq}
    \text{PFIC} = \frac{N_{p(x_{i}) > b}}{|X|}
\end{equation}
In Equation \ref{PFIC_eq}, $X$ is the design space, $x_{i}$ represents each candidate in the design space, $p(x_{i})$ represents the predicted performance of that candidate, $b$ is the performance of the best training data point, $|X|$ is the design space size, and $N_{\alpha}$ is the number of candidates satisfying condition $\alpha$. Therefore, the numerator of this equation represents the number of design space candidates predicted to be improvements over the best training data point, and the denominator represents the design space size. Equation \ref{PFIC_eq} assumes a maximization objective. A minimization objective can be achieved by subtracting the PFIC score from 1.
\par Fitting a linear model at each leaf node of the random forest was a critical choice for calculating the PFIC score. In typical random forest implementations, each leaf predicts a constant value equal to the average of the values of the training data at that leaf. Random forests that use this averaging approach are unable to predict values outside the range of the training data, so no candidates could be predicted to be an improvement over the best training data point. Other regression models that can extrapolate beyond the range of the training data to predict improvements could be paired with the PFIC score. Such regression models include Gaussian process regressors, support vector regressors, neural networks, kernel ridge regressors, and polynomial regressors. While these alternative algorithms were not investigated in this study, they would be of interest for future work.

\begin{figure}[ht]
\centering
\includegraphics[width=0.6\textwidth]{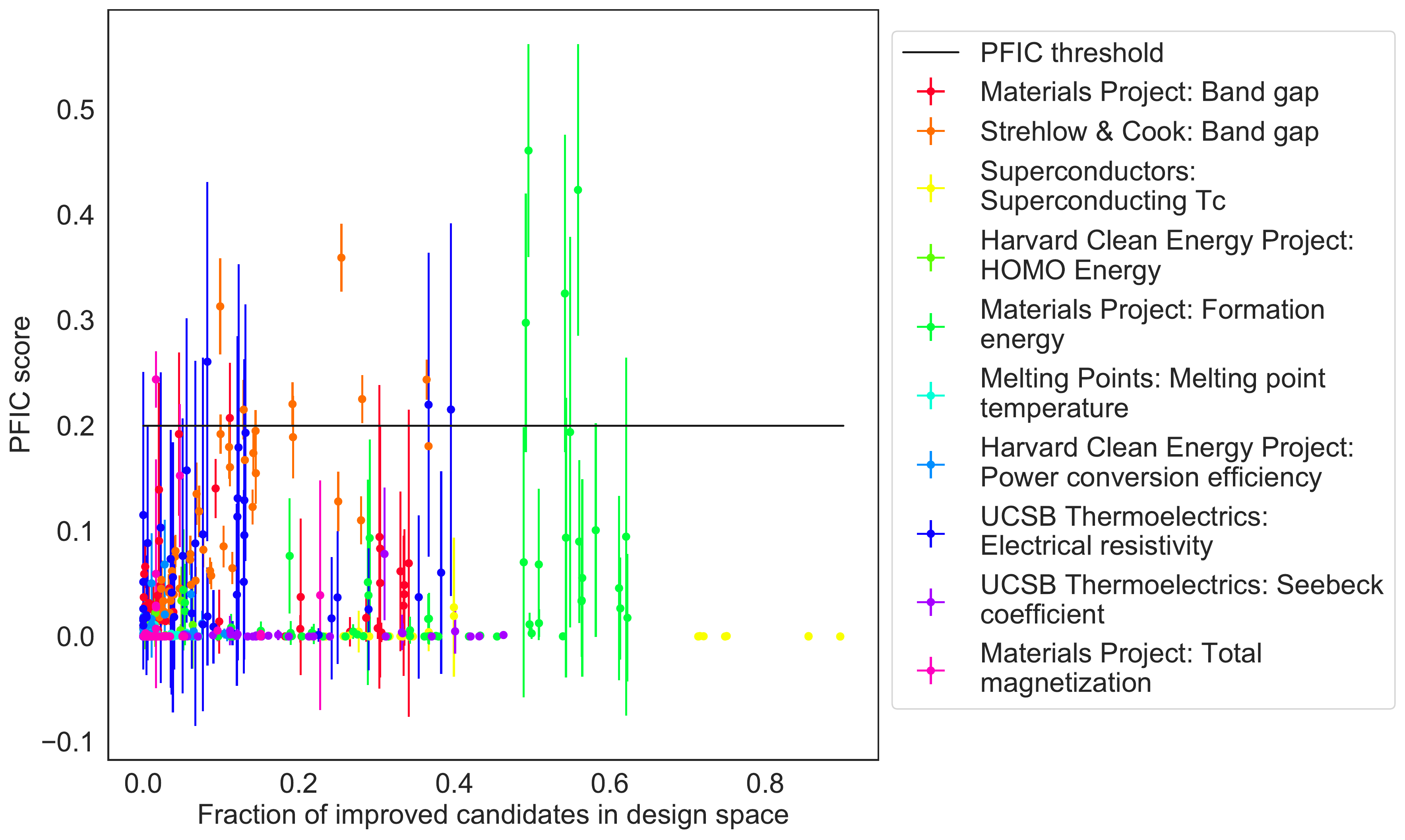}
\caption{PFIC score evaluated over benchmark design spaces versus the actual fraction of improved candidates in the design space. Each point represents a different design space and training data initialization, and the error bars on each point represents the variance in PFIC score over 20 identical data initializations. The black line indicates an example threshold value of $t_{PFIC} = 0.2$.}
\label{fig:pfic_score}
\end{figure}

\begin{figure}[ht]
\centering
\includegraphics[width=0.5\textwidth]{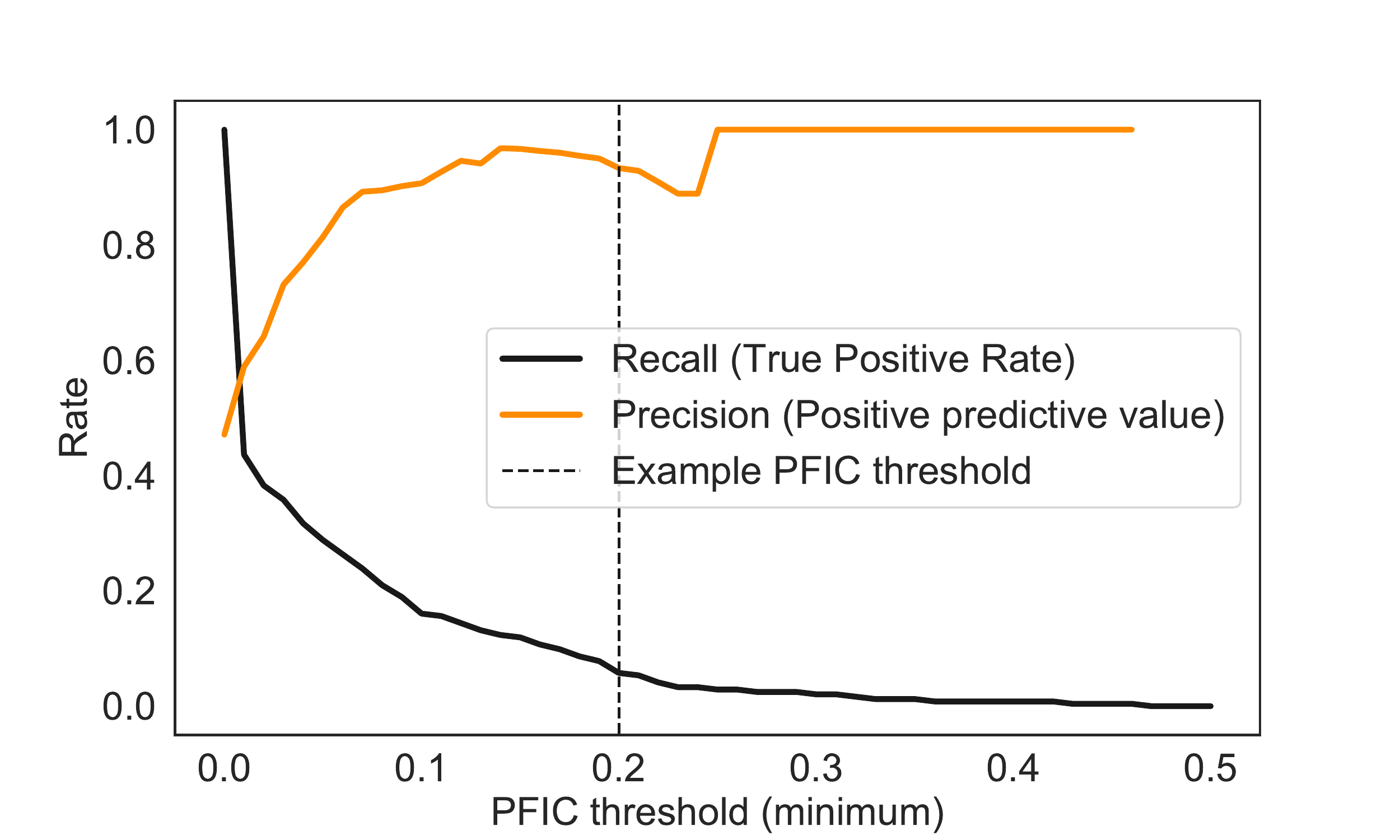}
\caption{Precision and recall of the PFIC score as a function of the PFIC threshold, $t_{PFIC}$, in identifying high quality design spaces, where high quality is defined as containing at least 4\% improved candidates in the design space. This 4\% corresponds to a baseline of 25 iterations until an improvement is found via random search over the design space. The dashed black line indicates an example threshold value of $t_{PFIC} = 0.2$.}
\label{fig:pfic_prec_rec}
\end{figure}

\begin{figure}[ht]
\centering
\includegraphics[width=0.4\textwidth]{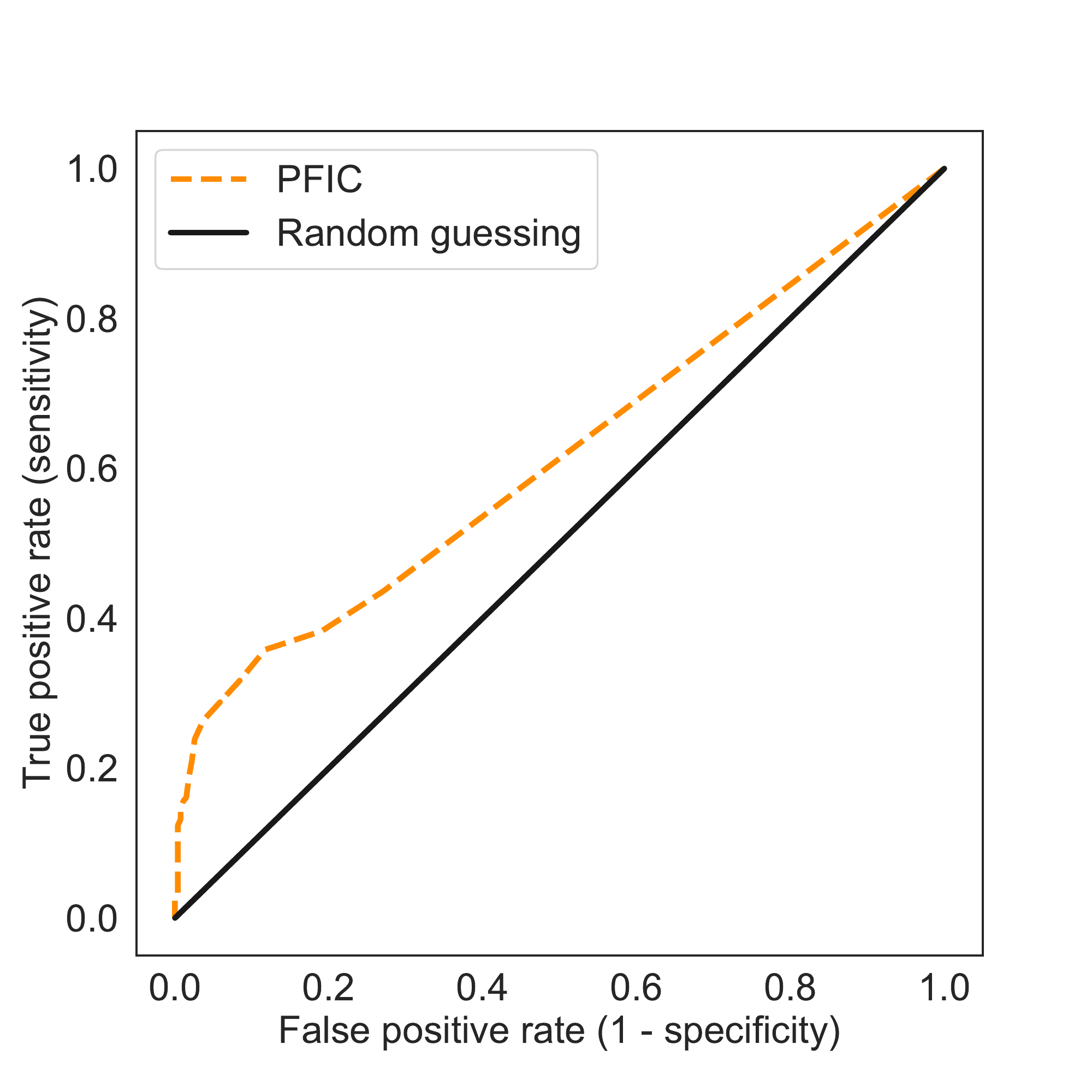}
\caption{ROC curve for PFIC score in identifying design spaces with fraction of improved candidates greater than 4\%. The AUC score for this curve is 0.62.}
\label{fig:pfic_roc}
\end{figure}

\par The PFIC score was tested on 10 different benchmarks from Table \ref{table:benchmarks}. The results of these tests are shown in Figure \ref{fig:pfic_score}. This figure shows the PFIC score as a function of the FIC, the true design space quality, across the benchmark datasets. Again, each point represents a separate data initialization, and the error bars represent the standard deviation over 20 trials of the same data initialization caused by stochastic model training. These results demonstrate that the PFIC score can be used to surface design spaces with high FIC at the outset of a project. In other words, the PFIC score could be used to highlight sequential learning projects that are likely to succeed because the haystack is full of needles. While this score does not correctly surface all high quality design spaces (i.e. there are some false negatives), it is still able to identify many high quality design spaces.
\par Figure \ref{fig:pfic_prec_rec} shows the precision and recall of the PFIC score in identifying high quality design spaces as a function of the PFIC threshold, $t_{PFIC}$. In this figure, we defined a high quality design space as containing at least 4\% improved candidates in the design space. This 4\% corresponds to a baseline of 25 iterations until an improvement is found via random search over the design space, which is within the typical range for sequential learning applications \cite{granda2018controlling, brandt2017rapid}. Precision is given by $\frac{TP}{TP+FP}$ and recall is given by $\frac{TP}{TP+FN}$, where $TP$ represents the true positives, or the design spaces correctly categorized as high quality, $FP$ represents the false positives, or the design spaces incorrectly categorized as high quality, and $FN$ represents the false negatives, or the high quality design spaces incorrectly categorized as low quality. Therefore, the precision represents the fraction of the design spaces identified as high quality by the PFIC score that are actually high quality, while recall represents the fraction of all high quality design spaces that were surfaced as high quality by the PFIC score. Figure \ref{fig:pfic_prec_rec} shows that as $t_{PFIC}$ increases, the precision improves while recall worsens. The notch in precision between $t_{PFIC}=0.2$ and $t_{PFIC}=0.3$ is caused by one sequential learning trial with $FIC<0.04$ and $PFIC>0.2$. Between $t_{PFIC}=0.2$ and $t_{PFIC}=0.25$, the precision continues to decrease, as other sequential learning trials with $FIC>0.04$ fall below $t_{PFIC}$. The precision shoots up to 1.0 once the one sequential learning trial with $t_{PFIC}<0.04$ finally falls below $t_{PFIC}$ at around $t_{PFIC}=0.25$.
\par Figure \ref{fig:pfic_roc} shows the ROC curve for the PFIC score in identifying high quality design spaces. The ROC curve plots the recall against the false positive rate, which is given by $\frac{FP}{TN+FP}$, where $FP$ represents the false positives, or the design spaces incorrectly categorized as high quality, and $TN$ represents the true negatives, or the low design spaces correctly categorized as low quality. The AUC score for this curve is 0.62. A perfect classifier has an AUC score of 1.0, while a classifier that is not capable of distinguishing between the high and low quality design spaces has an AUC score of 0.5. The PFIC score can thus be used to surface materials discovery problems with high quality design spaces.

\subsubsection*{Cumulative Maximum Likelihood of Improvement (CMLI) score}
\par The CMLI score is defined as the predicted probability that at least one candidate, out of the top $n$ candidates in the design space with the highest likelihoods of improvement, performs better than the best training data point, where $n$ is tunable by the experimenter. Calculating this score required using a machine learning model to predict the performance and estimate the uncertainty for each design space candidate.
\par This study used a random forest with constant-value leaves as the underlying machine learning algorithm \cite{lolo}. The uncertainty estimates were calculated using a combination of jackknife-based methods and an explicit bias model \cite{Ling2017,efron2012model}. However, other algorithms such as Gaussian process regressors could also be used to compute the CMLI score.
\par First, the top $n$ candidates with the highest likelihoods of improvement must be identified. The likelihood of improvement $L$ of a given design space candidate $x_{i}$ is given as: 
\begin{equation} \label{Likelihood_L_eq}
    L(x_{i}) = \int_{b}^{\infty} N(p(x_{i}), \sigma(x_{i}))dx
\end{equation}
In Equation \ref{Likelihood_L_eq}, $N(\mu, \sigma)$ is a normal distribution, with mean $\mu$  and standard deviation $\sigma$. The mean  $\mu$ is given by the predicted performance of candidate $x_{i}$ by the machine learning model, and the standard deviation $\sigma$ is given by the estimated uncertainty. Therefore, this integral represents the probability that design candidate $x_{i}$ is an improvement over the best training data point. Equation \ref{Likelihood_L_eq} assumes a maximization objective. Calculating the likelihood of improvement $L$ for a minimization objective would require changing the integration limits to $-\infty$ and $b$. While normally-distributed uncertainties were used in this study, $L$ could be readily computed for other uncertainty distributions as long as they are integrable. 
\par After the top $n$ candidates are identified, the CMLI score can be computed by the following equation: 
\begin{equation} \label{CMLI_eq}
    \text{CMLI} = 1 − \prod_{i=1}^{n} (1 − L(x_{i}))
\end{equation}
\par Equation \ref{CMLI_eq} takes the product of the likelihoods that candidates $x_{i}$ are \emph{not} an improvement, $1-L(x_{i})$. This product is over the top $n$ candidates from the design space with the highest likelihoods $L$ of performing better than the best training data point. This equation assumes that these likelihoods are independent for the top $n$ candidates. Therefore, this score assesses the likelihood that at least one of these top $n$ candidates is an improvement. In many cases, the model predictions are not independent, so the CMLI score has a tendency to overestimate the true design space quality. As a result, using $n = |X|$ results in CMLI scores near unity for large design spaces. This study uses CMLI score with $n$ at a value of 10. However, we have tested other values of $n$ which did not substantially affect the benchmarked test results. 

\begin{figure}[ht]
\centering
\includegraphics[width=0.6\textwidth]{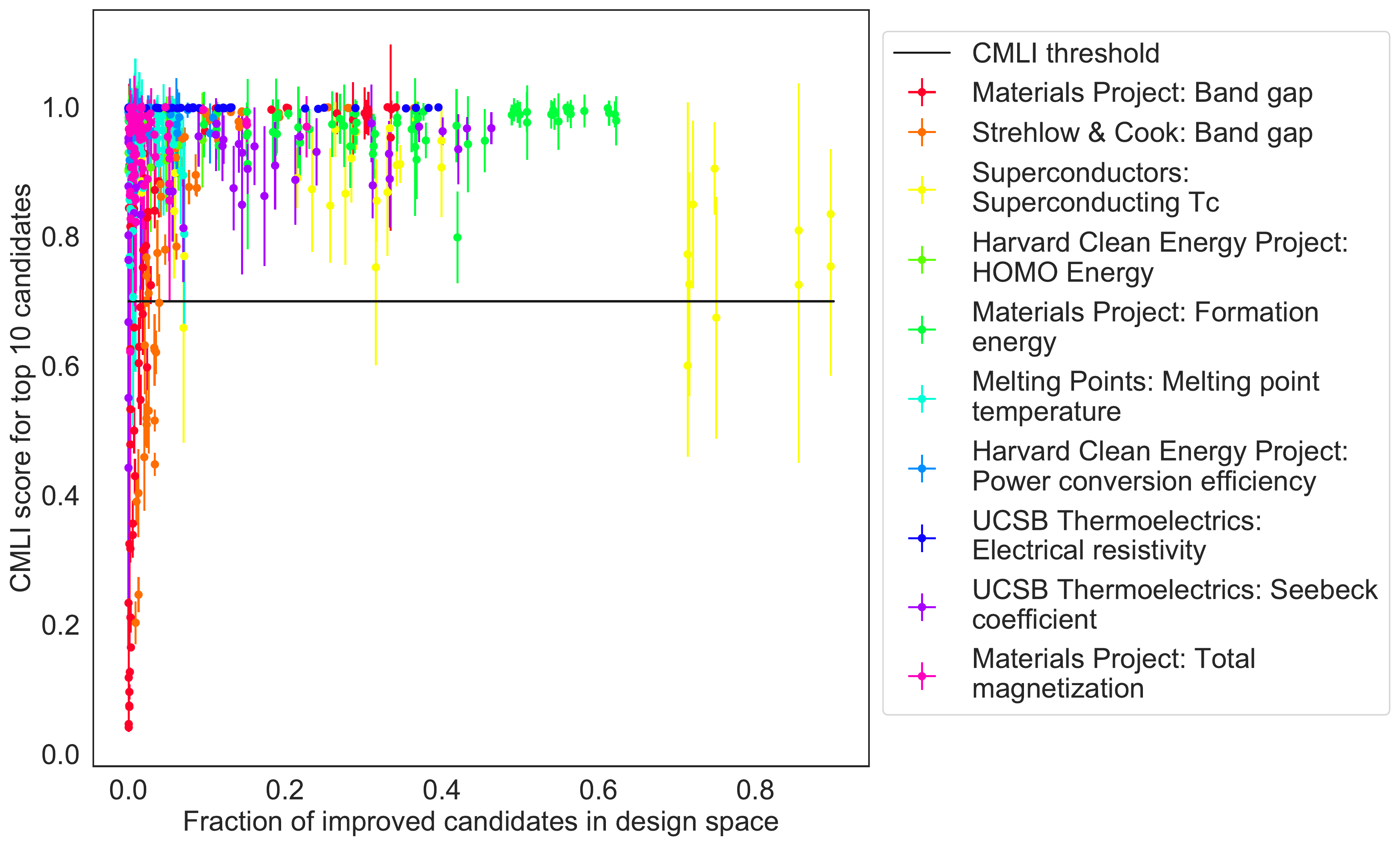}
\caption{CMLI score for top-10 candidates evaluated over benchmark design spaces versus the actual fraction of improved candidates in the design space. Each point represents a different design space and training data initialization, and the error bars on each point represents the variance in CMLI score over 20 identical data initializations. The black line indicates an example threshold value of $t_{CMLI} = 0.7$.}
\label{fig:cmli_score}
\end{figure}

\par The CMLI score was tested on 10 benchmark datasets from Table \ref{table:benchmarks}. The results from these tests are shown in Figure \ref{fig:cmli_score}. This figure shows the CMLI score versus the actual fraction of improved candidates across these benchmark datasets. This figure shows that the benchmark cases with low CMLI scores generally correspond to lower FICs, and that the CMLI score can be used to flag low quality design spaces at the outset of a project. Therefore, this score can be used to filter out sequential learning projects that may be difficult due to scarcity of needles in the haystack. While this score is not able to flag all low quality design spaces, it is able to identify some design spaces with low FIC.

\begin{figure}[ht]
\centering
\includegraphics[width=0.5\textwidth]{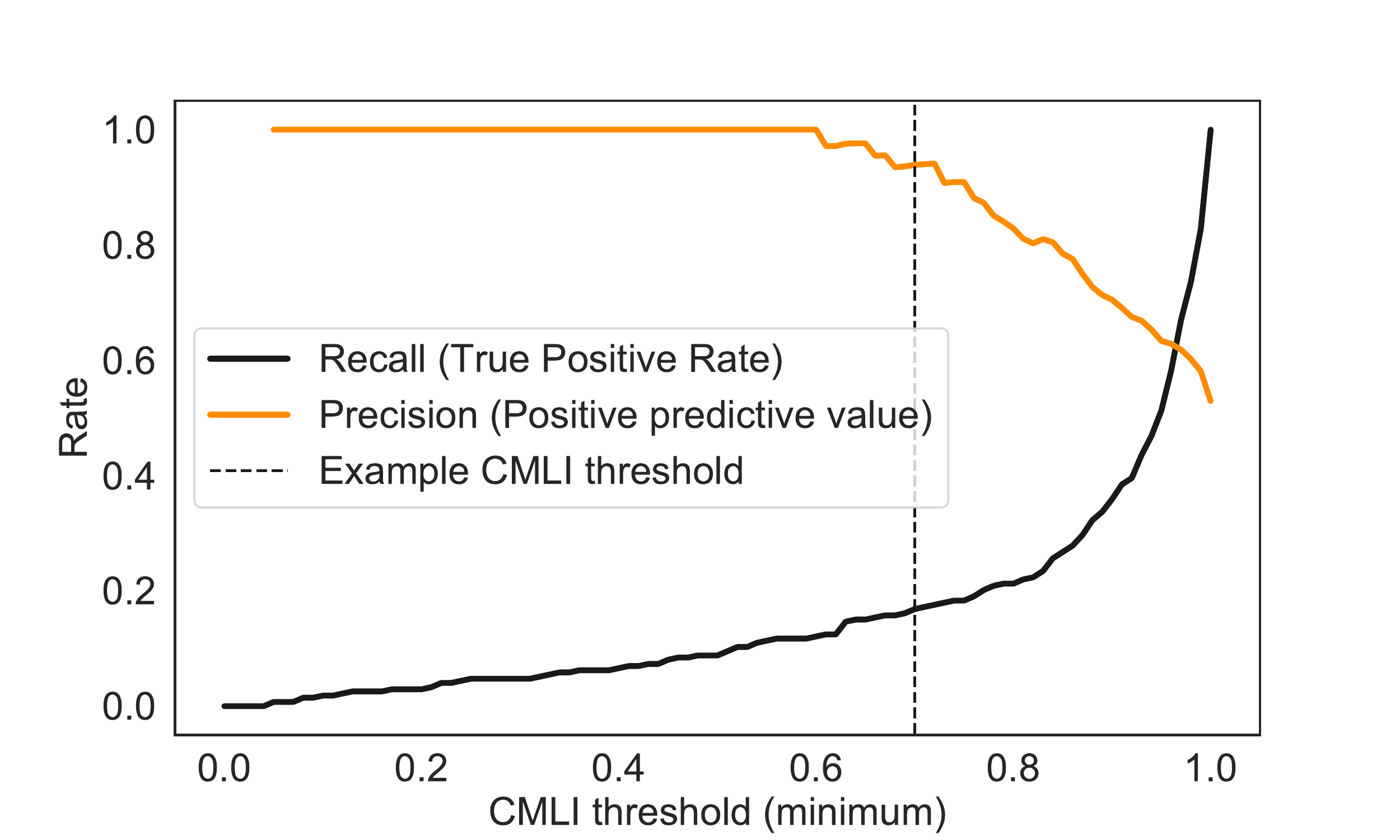}
\caption{Precision and recall of the CMLI score as a function of the CMLI threshold, $t_{CMLI}$, in identifying low quality design spaces, where low quality is defined as containing less than 4\% improved candidates in the design space. This 4\% corresponds to a baseline of 25 iterations until an improvement is found via random search over the design space. The dashed black line indicates an example threshold value of $t_{CMLI} = 0.7$.}
\label{fig:cmli_precision_recall}
\end{figure}

\begin{figure}[ht]
\centering
\includegraphics[width=0.4\textwidth]{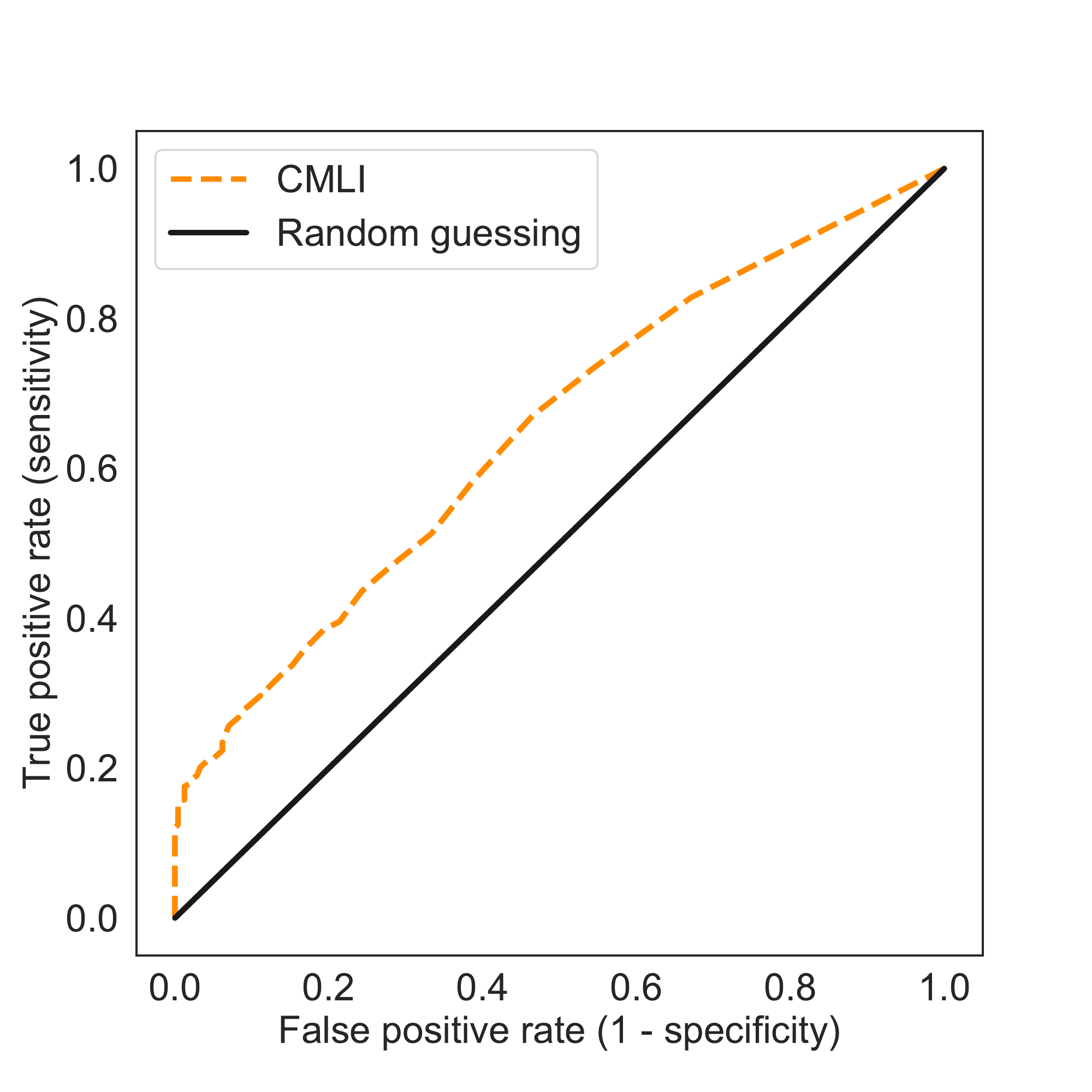}
\caption{ROC curve of CMLI score in identifying design spaces with fraction of improved candidates less than 4\%. The AUC score for this curve is 0.65.}
\label{fig:cmli_roc}
\end{figure}

\par Figure \ref{fig:cmli_precision_recall} shows the precision and recall of the CMLI score in identifying low quality design spaces as a function of the CMLI threshold, $t_{CMLI}$. In this figure, we defined a low quality design space as containing less than 4\% improved candidates in the design space. For the CMLI score identifying low quality design spaces, the precision represents the fraction of design spaces identified as low quality by the CMLI score that are actually low quality, while the recall represents the fraction of all low quality design spaces that were surfaced as low quality by the CMLI score. Figure \ref{fig:cmli_roc} shows the ROC curve for the CMLI score in identifying low quality design spaces. The AUC score for this curve is 0.65, where a perfect classifier has an AUC score of 1.0, while a classifier that is not capable of distinguishing between the high and low quality design spaces has an AUC score of 0.5. The CMLI score can thus be used to flag materials discovery problems with low quality design spaces.

\subsection*{Design space evaluation system}
\par We have investigated two metrics, the PFIC and CMLI scores, that can identify high quality and low quality design spaces respectively. Therefore, these two can be combined into a system for assessing design space quality. In the below description of our design space evaluation system, we refer to ML Models A and B. Model A must be capable of estimating uncertainty, and Model B must be able to predict output values that extrapolate outside the range of the training data, in order to predict improved design space candidate performance. In this work, we used two distinct models, but a single algorithm could potentially satisfy the requirements for both Models A and B. 
\begin{enumerate}
    \item Build a machine learning Model A with uncertainty estimates on the training data
    \item Use Model A to make predictions and uncertainty estimates for each design space candidate
    \item Use the predictions and uncertainty estimates from Model A to calculate the CMLI score for the top $n$ candidates
    \item Build a machine learning Model B on the training data
    \item Use this Model B to make predictions for each design candidate
    \item Use the predictions from Model B to evaluate the PFIC score
    \item Set thresholds $t_{CMLI}$ and $t_{PFIC}$ based on desired precision and recall
    \item Use the predictive metrics to assess design space quality
\end{enumerate}

\begin{figure}[ht]
\centering
\includegraphics[width=0.7\textwidth]{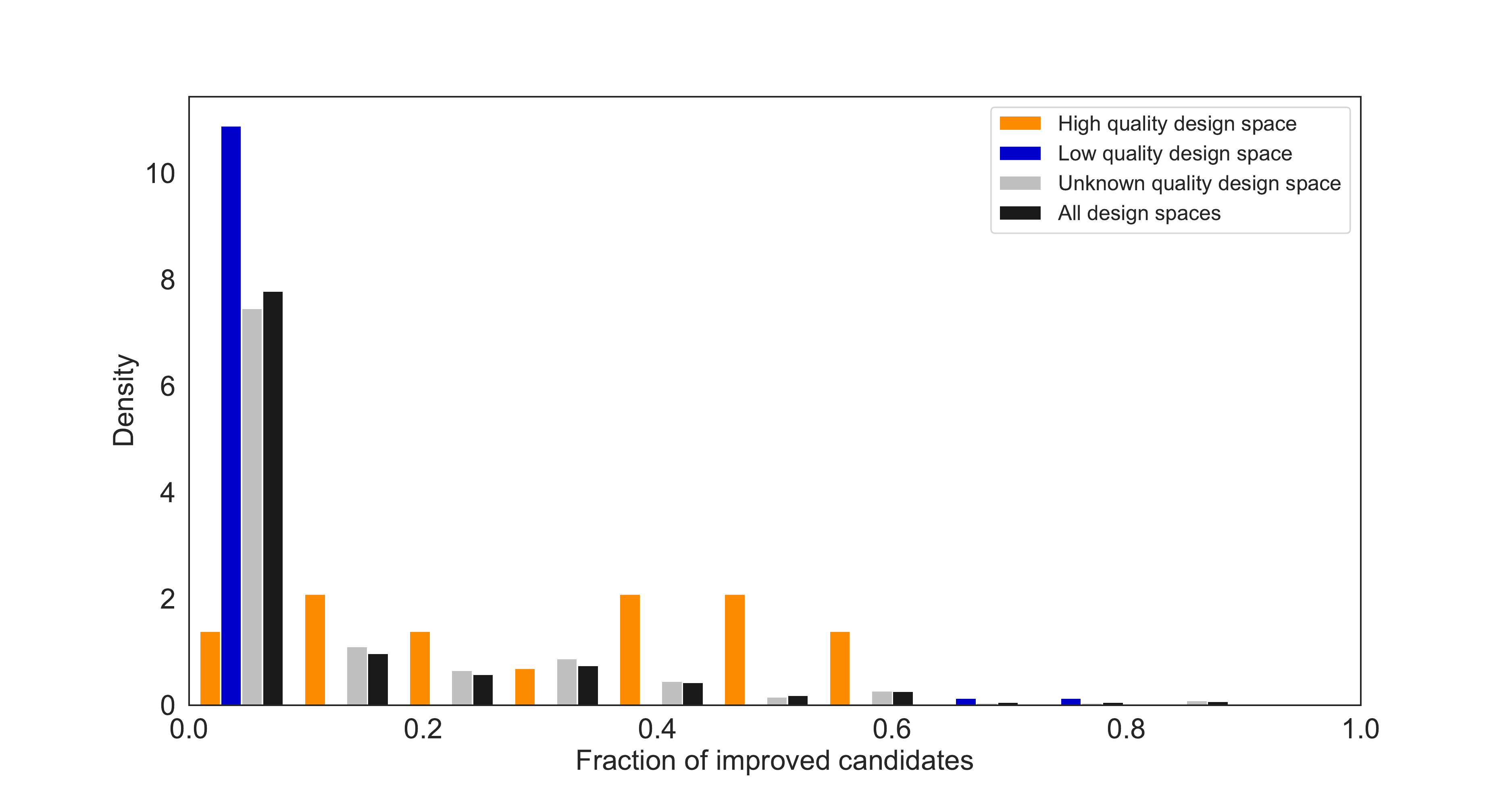}
\caption{Distribution of design space qualities as classified by design space evaluation system with $t_{PFIC} = 0.2$ and $t_{CMLI} = 0.7$. The y-axis shows the normalized count of design spaces such that the area sums to 1. }
\label{fig:evaluation_system_dist}
\end{figure}

Figure \ref{fig:evaluation_system_dist} shows the distribution of design spaces as classified by this design space evaluation system with example thresholds of $t_{PFIC} = 0.2$ and $t_{CMLI} = 0.7$. The x-axis is the fraction of improved candidates in the design space, and the y-axis is the density of the design spaces. The black bars show the overall distribution of design spaces. The orange bars represent the design spaces that were classified as high quality by the design space evaluation system, the blue bars represent design spaces classified as low quality, and the grey bars represent design spaces that were classified as unknown quality. 
\par In this example, we have defined high quality design spaces as having greater than 4\% of improved candidates in the design space, and low quality design spaces as having less than 4\% of improved candidates. Again, this 4\% corresponds to a baseline of 25 iterations until an improvement is found via random search over the design space, which is within the typical range for sequential learning applications \cite{granda2018controlling, brandt2017rapid}. Ideally, all design spaces with FIC $>0.04$ should be classified as high quality, those with FIC $<0.04$ should be classified as low quality, and there should not be any unknown quality design spaces. The distribution of all design spaces in black show that most design spaces have less than 10\% of improved candidates. This mirrors real materials development projects; very few projects have design spaces with many improvements, with the exception of nascent projects where not many materials have been explored. Additionally, the distribution of unknown quality design spaces closely follows the distribution of all design spaces. While the design space evaluation system is not perfect, Figure \ref{fig:evaluation_system_dist} shows that the design spaces classified as low quality are generally below 0.04 on the x-axis, while the design spaces classified as high quality are generally above 0.04 on the x-axis. Therefore, this system is useful in distinguishing between low and high quality design spaces. 
\par With these example thresholds, we are able to identify high quality design spaces with a precision of 0.94 and recall of 0.06. This means that 94\% of the design spaces identified as high quality by the design space evaluation system are actually high quality. Additionally, this means that we are able correctly surface 6\% of all the high quality design spaces. Meanwhile, we are able to flag low quality design spaces with a precision of 0.96 and a recall of 0.23. This means that 96\% of the design spaces flagged as low quality are actually low quality, and that we are able to correctly identify 23\% of all low quality design spaces. In this example, the design space evaluation system is lossy - many design spaces are categorized as unknown quality. However, we can be fairly confident in design spaces that are categorized as high or low quality. Additionally, the thresholds for the predictive metrics can be tuned to achieve the desired precision and recall scores for both high quality and low quality design spaces. 
\par This design space evaluation system uses both the PFIC and CMLI scores to identify high quality and low quality design spaces. Given the strong impact of design space quality on the difficulty of the overarching sequential learning project, this design space evaluation system is a useful tool for assessing the probability of success for any given materials development project.

\section*{CONCLUSION}
\par While many previous studies have explored the promise of machine learning methods for accelerating materials development, the use of machine learning to assess the difficulty of a materials development project \textit{a priori} represents a new contribution. However, it would be extremely valuable to know how hard a materials development project would be at the outset.  This information could be used to determine which projects to invest in and how to allocate resources across a research and development portfolio.  For groups investing in multiple projects simultaneously, information on project difficulty could be used to balance the project portfolio between high-risk and low-risk projects. 
\par In this work, we demonstrated the importance of design space quality on materials discovery success, introduced a new data initialization method to reflect in-lab materials discovery, and finally, defined novel predictive metrics to determine the quality of a design space at the outset of a materials discovery project. 
\par After evaluating these design space metrics across a variety of computational and experimental materials datasets, we identify two scores - PFIC and CMLI - which correlate with the true design space quality. These two metrics are then combined into a high-precision, model-agnostic design space evaluation system. Our work is thus a first step towards determining, \emph{a priori}, the difficulty of a materials development project.  While all the trials run in this study used a simulated sequential learning workflow, these predictive metrics could also be used in more traditional development settings where experimental test order was determined by a scientist instead of a machine learning algorithm.  In all cases, the fraction of improved candidates in the design space will be strongly related to the difficulty of a materials development project.   
\par Future work may include testing additional predictive metrics to reveal further insights on design space quality. Future work may also include correlating predictive design space metrics directly to sequential learning success (iterations until an improvement is found) rather than the FIC, in order gain further insight on how these predictive metrics can be used to prioritize sequential learning projects. These metrics may also be calculated at every iteration of sequential learning to analyze how these metrics change when new data points are added to the training data. Additionally, further experiments may be conducted to determine how model accuracy, algorithmic choice, and uncertainty quantification approach affect the accuracy of our predictive design space metrics. Finally, the metrics discussed in this work concern the optimization of single materials properties, and future work may address the challenge of multi-objective optimization.

\bibliography{sample}

\section*{Acknowledgements}
The authors would like to thank Astha Garg for helpful discussions in brainstorming sequential learning metrics. The authors would also like to thank Suzanne Petryk for assistance in developing the framework for sequential learning simulations.

\section*{Author contributions statement}
Y.K. and E.A. ran the sequential learning simulations and analyzed the results. E.A. created the framework for running these sequential learning simulations. Y.K., E.K., E.A., B.M., and J.L. designed the study and wrote the manuscript.

\section*{Additional information}

\textbf{Competing Interests} The authors declare that a US patent application has been filed relating to the use of a design space evaluation system for materials development. This patent application has been filed by Citrine Informatics, with all co-authors listed as co-inventors. At the time of article submission, the patent application is pending examination and has been assigned the application number 16/568,701. 

\end{document}